\documentclass[12pt]{article}
\usepackage{nature}
\usepackage{amssymb}
\usepackage{graphics}
\begin{document}
\baselineskip 24pt
\bf
\noindent
{\huge 
Structure in the early afterglow lightcurve of the $\gamma$-ray burst of
29 March 2003}\\

\noindent
Makoto Uemura$^*$, Taichi Kato$^*$, Ryoko Ishioka$^*$, Hitoshi
Yamaoka$^\dag$, Berto Monard$^\ddag$, Daisaku Nogami$^\S$, Hiroyuki
Maehara$^\|$, Atsushi Sugie$^\natural$, and Susumu Takahashi$^\natural$\\
$^*$ Department of Astronomy, Kyoto University, Kyoto 606-8502, Japan\\
$^\dag$  Faculty of Science, Kyushu University, Fukuoka 810-8560\\
$^\ddag$ Bronberg Observatory, PO Box 11426, Tiegerpoort 0056, South
Africa\\
$^\S$ Hida Observatory, Kyoto University, Kamitakara, Gifu 506-1314\\
$^\|$ VSOLJ, Namiki 1-13-4, Kawaguchi, Saitama 332-0034\\
$^\natural$ Dynic Astronomical Observatory, Taga 283-1, Taga-chou,
Inugami, Shiga, 522-0341 ,Japan\\
\begin{abstract}
\bf
 Gamma-ray bursts (GRBs) are energetic explosions that for 0.01--100 s
 are the brightest $\gamma$-ray sources in the sky.\cite{wij97GRB970228}
 \cite{wax97afterglow}  Observations of the early evolution of
 afterglows we expected to provide clues about the nature of the bursts,
 but their rapid fading has hampered such studies; some recent rapid
 localizations\cite{fox03GRB021004} \cite{wei03GRB021211}
 \cite{fox03GRB021211} of bursts have improved the situation.  Here we
 report on an early detection of the very bright afterglow of the burst
 of 29 March 2003 (GRB030329).  Our data show that, even early in the
 aferglow phase, the light curve shows unexpectedly complicated
 structures superimposed on the fading background. 
\end{abstract}
\rm
GRB030329 was detected by the HETE-2 satellite on 29 March 2003
at 11:37:14.67 UT.\cite{van03gcn1997}  A prompt identification of
its burst position, which was notified 0.051 d after the burst,
enabled observers to make observations during an early phase of the
afterglow of the GRB.  A new very bright optical source of magnitude 12 
was discovered within the error circle of the GRB position, and then
continuously monitored by a number of observers.\cite{pet03gcn1985}
\cite{tor03gcn1986} \cite{uem03gcn1989}  Optical spectroscopic
observations determined its redshift to be $z=0.168$, and using this,
its isotropic released energy was calculated to be $E_{\rm iso}=8\times
10^{51}\;{\rm erg\, s^{-1}}$ (30--400 keV).\cite{gre03gcn2020}
\cite{cal03gcn2053}  In the 30-years of studying GRBs, this GRB
afterglow is the closest except for the suspected supernova
associated with GRB980425.\cite{tin98GRB980425}
 
We initiated time-series CCD photometric observations on 29 March 2003 at
12:53:41 UT, about 0.053 days after the burst, at Kyoto University, with 25-
and 30-cm telescopes.  After the observation at Kyoto,
observations at Saitama, the Dynic Astronomical Observatory, and the
Bronberg Observatory started with 20-cm, 60-cm, and 30-cm telescopes,
respectively.  Follow-up observations were performed on 30 March 30, 3,
5, and 6 April at Kyoto, on 30 March at Hida, and on 30 and 31 March at
Bronberg.  A dark-current image was subtracted from obtained CCD images,
and then flat-fielding was performed.  We calculated the magnitudes of
the afterglow with neighbour comparison stars (GSC1434.239, GSC1434.129,
and GCS1434.192), whose constancy was better than 0.04 mag during our
observation.  Our unfiltered CCD observation yields a magnitude system
near that of the $R_{\rm c}$-system, since the sensitivity peak of the
camera is near to the peak of the $R_{\rm c}$-system, and the spectra of early
afterglows have a smooth continuum without strong emission lines or
absorption edges in the optical range.  The difference between our
unfiltered CCD and the $R_{\rm c}$-system is less than 0.02 mag when the
spectrum is described with a simple power law ($f(\nu)\propto \nu^p$)
with index, $-2.0<p<0.3$, as expected in GRB optical
afterglows.\cite{sar98grb} 

Our obtained light curve is shown in Fig. 1.  In Fig. 1a, the
abscissa and the ordinate denote the time from the burst in days and the
$R_{\rm c}$-magnitude, respectively.  To date, the light curve of
afterglows, in particular during the early phase of $\lesssim 0.1$ days, 
has been sampled only sparsely, which has been insufficient to
observationally verify theoretical models.  As can be seen in this
figure, we succeeded in obtaining a completely continuous light curve
from 0.053 to 0.500 days, or for 11 hours in GRB030329, which reveals
the presence of unambiguous detailed structures in the light curve of
early afterglows. 

The light curve of the GRB030329 afterglow exhibits clear, and repeating
deviations from a canonical power-law model.  The dotted line in Fig. 1a
is the best fitted power-law model using our light
curve and the GRB Coordinates Network data between 0.5 and 10 days.  The
average decay index, 
$\alpha$ ($f\propto t^\alpha$) is calculated to be $-1.115\pm 0.006$.
This is a standard value for early afterglows of GRBs.  As can be seen
in the figure, the fading of the afterglow cannot be described only by
this simple model.  The residual from the single power-law model is
shown in Fig. 1b.  During the first day, the light
curve can be described with four segments, that is, two sets of ascending
and descending branches of bumps.  Their transitions were not smooth,
but occurred in short timespans.  A relatively rapid fading presumably
follows these bumps.  At the end of the first day, the afterglow ceased
fading, and then, experienced a rebrightening.  While they were
sampled only partially, the light curve shows that the bumps after 1 d
had a rising time-scale of 0.1 day.  It is interesting to note that an
early bump around 0.4 day in Fig. 1 had a timescale of the same order as the
later bumps.  This similarity might imply that these large (amplitude
$\sim 0.4$--$0.5$ mag) bumps were of the same nature. 

The afterglow of GRBs is now widely believed to be synchrotron emission
from a forward shock region in an expanding jet colliding with ambient
medium, whereas the GRB itself is from an internal shock region between
shells.\cite{wij97GRB970228} \cite{wax97afterglow} 
\cite{sar99prediction}  In a number of afterglows, it is well known that
light curves show a sudden increase of the fading rate due to the jet
geometry, typically a few days after the burst.\cite{rho99lightcurve}
\cite{sta99GRB990510}  In the case of GRB030329, this jet break is
expected to appear later than 60 days from its isotropic energy and
redshift.\cite{mal02gcn1607} \cite{fra01standard}  This indicates that
the jet break is expected not to appear in Fig. 1, and hence, it is
difficult to explain the variation in Fig. 1 with the jet break.  As
well as the jet break, another well-known phenomenon, the supernova bump
is evidently difficult to explain the multiple
modulations.\cite{rei99GRB970228}  In GRB030329, the feature of a
supernova spectrum appeared 7 days after the burst.\cite{sta03SN2003dh}
The optical flux before 7 days was hence dominated by the afterglow
emission itself.  

In some GRB afterglows, similar deviations from a general fading
component have been reported.\cite{gal98GRB970508} \cite{pir98GRB970508}
\cite{gar00GRB000301C}  The recent object, GRB021004 is 
the first object exhibiting multiple fluctuations 1 day after the
burst.\cite{nak03GRB021004}  In GRB030329, we found that the large
amplitude bumps appeared even in the first day after the burst.  The
mechanism of these bumps has not been established yet, but the
continuous light curve of the bumps of GRB030329 should help us to
understand them. 


\begin{thebibliography}{10}

\bibitem{wij97GRB970228}
Wijers, R. A. M.~J., Rees, M.~J., \& Meszaros, P.
\newblock Shocked by GRB 970228: the afterglow of a cosmological fireball.
\newblock {\em \mnras}{ \bf 288}, L51-L56 (1997).

\bibitem{wax97afterglow}
Waxman, E.
\newblock Gamma-Ray--Burst Afterglow: Supporting the Cosmological Fireball
  Model, Constraining Parameters, and Making Predictions.
\newblock {\em \apj}{ \bf 485}, L5-L8 (1997).

\bibitem{fox03GRB021004}
Fox, D.~W. et al.
\newblock Early optical emission from the ¦Ã-ray burst of 4 October 2002.
\newblock {\em Nature}{ \bf 422}, 284-286 (2003).

\bibitem{wei03GRB021211}
Weidong, L., Alexei, F.~V., Ryan, C., \& Saurabh, J.
\newblock The Early Light Curve of the Optical Afterglow of GRB 021211.
\newblock {\em \apj}{ \bf 586}, L9-L12 (2003).

\bibitem{fox03GRB021211}
Fox, D.~W. et al.
\newblock Discovery of Early Optical Emission from GRB 021211.
\newblock {\em \apj}{ \bf 586}, L5-L8 (2003).

\bibitem{van03gcn1997}
Vanderspek, R. et al.
\newblock RB030329 (=H2652): A Long, Extremely Bright GRB Localized by the HETE
  WXM and SXC.
\newblock {\em GRB Circ. Netw.}{ \bf 1997} (2003).

\bibitem{pet03gcn1985}
Peterson, B.~A. \& Price, P.~A.
\newblock GRB 030329: Optical afterglow candidate.
\newblock {\em GRB Circ. Netw.}{ \bf 1985} (2003).

\bibitem{tor03gcn1986}
Torii, K.
\newblock GRB 030329: OT candidate.
\newblock {\em GRB Circ. Netw.}{ \bf 1986} (2003).

\bibitem{uem03gcn1989}
Uemura, M.
\newblock GRB 030329: Optical afterglow fading.
\newblock {\em GRB Circ. Netw.}{ \bf 1989} (2003).

\bibitem{gre03gcn2020}
Greiner, J. et al.
\newblock Redshift of GRB 030329.
\newblock {\em GRB Circ. Netw.}{ \bf 2020} (2003).

\bibitem{cal03gcn2053}
Caldwell, N., Garnavich, P., Holland, S., Matheson, T., \& Stanek, K.~Z.
\newblock GRB 030329, optical spectroscopy.
\newblock {\em GRB Circ. Netw.}{ \bf 2053} (2003).

\bibitem{tin98GRB980425}
Tinney, C. et al.
\newblock GRB 980425.
\newblock {\em IAU Circ.}{ \bf 6896} (1998).

\bibitem{sar98grb}
Sari, R., Piran, T., \& Narayan, R.
\newblock Spectra and Light Curves of Gamma-Ray Burst Afterglows.
\newblock {\em \apj}{ \bf 497}, L17-L20 (1998).

\bibitem{sar99prediction}
Sari, R. \& Piran, T.
\newblock Predictions for the Very Early Afterglow and the Optical Flash.
\newblock {\em \apj}{ \bf 520}, 641-649 (1999).

\bibitem{rho99lightcurve}
Rhoads, J.~E.
\newblock The Dynamics and Light Curves of Beamed Gamma-Ray Burst Afterglows.
\newblock {\em \apj}{ \bf 525}, 737-749 (1999).

\bibitem{sta99GRB990510}
Stanek, K.~Z., Garnavich, P.~M., Kaluzny, J., Pych, W., \& Thompson, I.
\newblock BVRI Observations of the Optical Afterglow of GRB 990510.
\newblock {\em \apj}{ \bf 522}, L39-L42 (1999).

\bibitem{mal02gcn1607}
Malesani, D. et al.
\newblock GRB021004: optical observations and predicted break time.
\newblock {\em GCN}{ \bf 1607} (2002).

\bibitem{fra01standard}
Frail, D.~A. et al.
\newblock Beaming in Gamma-Ray Bursts: Evidence for a Standard Energy
  Reservoir.
\newblock {\em \apj}{ \bf 562}, L55-L58 (2001).

\bibitem{rei99GRB970228}
Reichart, D.~E.
\newblock GRB 970228 Revisited: Evidence for a Supernova in the Light Curve and
  Late Spectral Energy Distribution of the Afterglow.
\newblock {\em \apj}{ \bf 521}, L111-L115 (1999).

\bibitem{sta03SN2003dh}
Stanek, K.~Z. et al.
\newblock Spectroscopic Discovery of the Supernova 2003dh Associated with GRB
  030329.
\newblock {\em \apj}{ \bf in the press}, (astro--ph/0304173) (2003).

\bibitem{gal98GRB970508}
Galama, T.~J. et al.
\newblock Optical Follow-Up of GRB 970508.
\newblock {\em \apj}{ \bf 497}, L13-L16 (1998).

\bibitem{pir98GRB970508}
Piro, L. et al.
\newblock Evidence for a late-time outburst of the X-ray afterglow of GB970508
  from BeppoSAX.
\newblock {\em \aap}{ \bf 331}, L41-L44 (1998).

\bibitem{gar00GRB000301C}
Garnavich, P.~M., Loeb, A., \& Stanek, K.~Z.
\newblock Resolving Gamma-Ray Burst 000301C with a Gravitational Microlens.
\newblock {\em \apj}{ \bf 544}, L11-L15 (2000).

\bibitem{nak03GRB021004}
Nakar, E., Piran, T., \& Granot, J.
\newblock Variability in GRB afterglows and GRB 021004.
\newblock {\em New Astronomy}{ \bf 8}, 495-505 (2003).

\bibitem{beu990510}
Beuermann, K. et al.
\newblock VLT observations of GRB 990510 and its environment.
\newblock {\em \aap}{ \bf 352}, L26-L30 (1999).

\end{thebibliography}

\noindent
Acknowledgements

This work is partly supported by a grant-in aid from the Japanese
Ministry of Education, Culture, Sports, Science and Technology.  Part of
this work is supported by a Research Fellowship of the Japan Society for
the Promotion of Science for Young Scientists (MU\ and RI).

\clearpage

\includegraphics{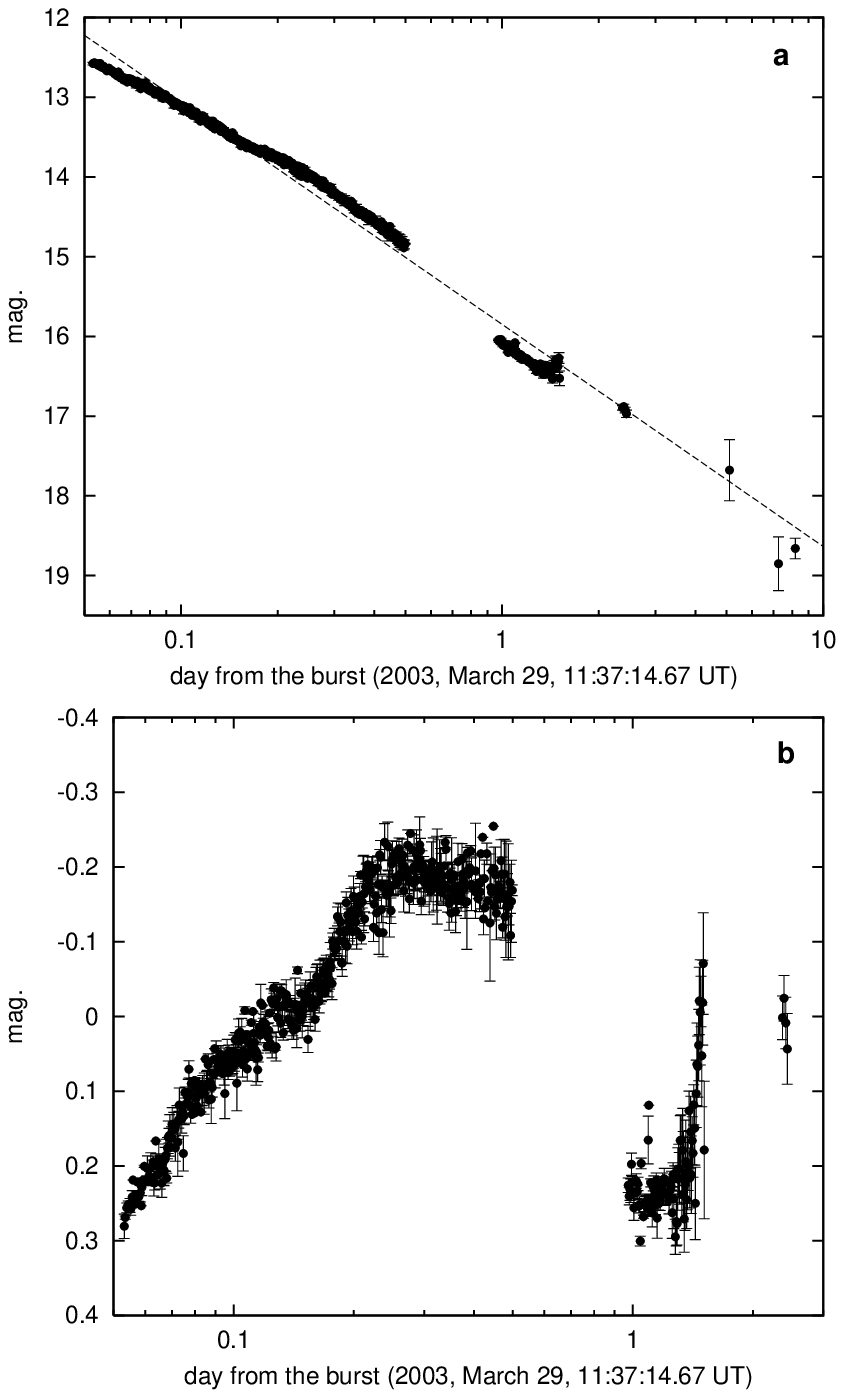}

\noindent
\textsf{Figure 1}. Light curves of the optical afterglow of GRB030329.
The abscissa denotes the time from the burst (29 March 2003,
11:37:14.67 UT in days, shown in logarithmic scale.  The ordinate
denotes the $R_{\rm c}$-magnitude in \textsf{a} and the residual
magnitude from a power-law decay in \textsf{b}.  The dotted line in 
\textsf{a} is the best-fitted power-law model, which was used to
calculate the residual magnitude in \textsf{b}.  
Our observations are indicated with filled circles with error bars,
indicating standard errors.  The exposure times of each frame were 30,
40, 60, 30-60and 45 s for Kyoto, Saitama, the Dynic Astronomical Observatory,
the Hida Observatory 
and the Bronberg observations, respectively.  The points in the figure
are binned points of the observations with $\Delta \log{t_{\rm day}} =
0.002$.  Using a smoothly broken power-law 
model ($f(t)=(f_1(t)^{-n}+f_2(t)^{-n})^{-1/n}$ with $f_i(t)=k_i
t^{-\alpha_i}$),\cite{beu990510} we determined decay indices ($\alpha_i$)
to be $0.74\pm 0.02\;(0.053<t<0.085)$, $0.95\pm 0.01\;(0.085<t0.163)$,
$0.65\pm 0.04\;(0.163<t<0.227)$, and $1.16\pm 0.01\;(0.227<t<0.492)$.
The break times are $0.085\pm 0.028$, $0.163\pm 0.060$, and $0.227\pm
0.043$.  Note that the 0.16-day break time was calculated with a broken
power-law model without the smoothness parameter.  The smoothness
parameters, $n$, are $94\pm51$ for the earlier break and $69\pm 89$ for
the later break, indicating rapid state transitions.

\end{document}